\documentclass{aa}
\usepackage{graphicx}
\usepackage[varg]{txfonts}
\usepackage{cases}
\usepackage{natbib}
\usepackage{amsmath}

\newcommand{\figref}[1]{Figure~\ref{#1}}
\newcommand{\secref}[1]{Section~\ref{#1}}

\begin{document}

\title{An ultra-long and quite thin coronal loop without significant expansion}

\author{Dong~Li\inst{1,2}, Ding~Yuan\inst{3}\thanks{Corresponding author}, Marcel~Goossens\inst{4}, Tom~Van~Doorsselaere\inst{4}, Wei~Su\inst{5}, Ya~Wang\inst{1}, Yang~Su\inst{1,6}, \& Zongjun~Ning\inst{1,6}}

\institute{Key Laboratory of Dark Matter and Space Astronomy, Purple Mountain Observatory, CAS, Nanjing 210033, PR China \email{lidong@pmo.ac.cn} \\
           \and State Key Laboratory of Space Weather, Chinese Academy of Sciences, Beijing 100190, PR China \\
           \and Institute of Space Science and Applied Technology, Harbin Institute of Technology, Shenzhen 518055, PR China \email{yuanding@hit.edu.cn} \\
           \and Centre for mathematical Plasma Astrophysics, Department of Mathematics, KU Leuven, Celestijnenlaan 200B, 3001 Leuven, Belgium \\
           \and MOE Key Laboratory of Fundamental Physical Quantities Measurements, School of Physics, Huazhong University of Science and Technology, Wuhan 430074, PR China \\
           \and School of Astronomy and Space Science, University of Science and Technology of China, Hefei 230026, PR China \\}
\date{Received; accepted}

\titlerunning{An ultra-long and quite thin coronal loop without significant expansion}
\authorrunning{Dong Li et al.}

\abstract {Coronal loops are the basic building blocks of the solar
corona, which are related to the mass supply and heating of solar
plasmas in the corona. However, their fundamental magnetic
structures are still not well understood. Most coronal loops
do not expand significantly, whereas the diverging magnetic field
would have an expansion factor of about 5-10 over one pressure scale
height.}{In this study, we investigate a unique coronal loop with
a roughly constant cross section, it is ultra long and quite thin.
A coronal loop model with magnetic helicity is presented to
explain the small expansion of the loop width.} {This coronal loop
was predominantly detectable in the 171~{\AA} channel of the
Atmospheric Imaging Assembly (AIA). Then, the local magnetic field
line was extrapolated by a Potential-Field-Source-Surface model.
Finally, the differential emission measure analysis made from six
AIA bandpasses was applied to obtain the thermal properties
of this loop.} {This coronal loop has a projected length of roughly
130~Mm, a width of about $1.5\pm0.5$~Mm and a lifetime of around
90~minutes. It follows an open magnetic field line. The cross
section expanded very little (i.e., 1.5$-$2.0) along the loop length
during its whole lifetime. This loop has a nearly constant
temperature at about $0.7\pm0.2$~MK, whereas its density exhibits
the typical structure of a stratified atmosphere.} {We use a thin
twisted flux tube theory to construct a model for this non-expanding
loop, and find that indeed with sufficient twist a coronal loop can
attain equilibrium. However, we can not rule out other possibilities
such as footpoint heating by small-scale reconnection,
elevated scale height by a steady flow along the loop etc.}


\maketitle

\section{Introduction}
Coronal loops are the basic structures in the solar corona. They can
be detected everywhere on the Sun, such as the quiet region, the
active region, or the solar limb, and their size scale could be
ranging from sub-Megameter to hundreds of Megameters in the lower
corona. These coronal loops often confine plasmas at the temperature
of Mega-Kelvin, so they are prominently detectable in the extreme
ultraviolet (EUV) and X-ray bandpasses \citep{Bray91,Reale14}.
Moreover, the plasmas contained in a coronal loop may be either
isothermal \citep[e.g.,][]{Delz03,Tripathi09,Gupta19} or
multithermal \cite[e.g.,][]{Schmelz06,Kucera19} along the line of
sight. In the corona, full-ionized plasma is frozen-in the magnetic
field line. So the plasma properties are normally uniform along the
loops, strong inhomogeneity are usually detected across the loops.
By comparisons of the coronal imaging observations in EUV or X-ray
channels, together with the extrapolated field lines derived from
the photospheric magnetogram, the coronal loop was found to
generally follow the magnetic field line \citep{Poletto75,Feng07}.
That is, a closed coronal loop is usually consist of a loop apex and
two footpoints rooted in two opposite polarities
\citep[e.g.,][]{Watko00,Peter12}, while an open coronal loop
connects to one apparent polarity at the solar surface and extends
radially into the heliosphere magnetic field
\citep[e.g.,][]{Gupta19}. Previous studies also suggested that the
temperature variation along a coronal loop is highly sensitive to
the heating mechanism \citep{Priest98,Warren08}. Therefore, to study
the coronal loop in the complex magnetic environment could help us
to better understand the fundamental problem in solar physics, i.e.,
coronal heating \citep[e.g.,][]{Klimchuk00,Peter12,Li15,Goddard17}.

The coronal loop is expected to expand with height, since the
coronal magnetic field is found to diverge strongly with the height
from the solar surface into the corona \citep{Lionello13,Chen14}.
The expansion of coronal loop could be discovered on the active
region \citep[e.g.,][]{Malanushenko13} or solar limb
\citep[e.g.,][]{Gupta19}. However, most of coronal loops observed in
X-ray and EUV images are found to have roughly uniform widths in the
plane of the sky, without significant expansions along their loop
lengths, or only exhibit a small expansion from footpoints to the
loop apex
\citep[e.g.,][]{Golub90,Klimchuk92,Klimchuk00,Watko00,Lopez06,Brooks07,Kucera19}.
The formation and appearance of these loops in the complex magnetic
environment of the corona provides a pivotal test for a model of the
coronal heating process \citep{Klimchuk00,Petrie06,Peter12}. On the
other hand, the loop cross section carries the information of
magnetic fields and the spatial distribution of corona heating, and
the lower limit of the loop width is of fundamental importance to
modern instrumentation, because it defines the spatial resolution of
a space-borne or ground-based telescope
\citep{Peter13,Aschwanden17}. Moreover, the loop width variation is
a proxy of the inter-coupling of plasma dynamics and magnetic
fields, so it is believed to play a key role in coronal heating
\citep[e.g.,][]{Vesecky79,McTiernan90,Mikic13,Chastain17}.
\cite{Aschwanden17} find that the loop widths are marginally
resolved in AIA images but are fully resolved in Hi-C images, their
model predicts a most frequent value at about 0.55~Mm.

The contradiction between the observed coronal loop with a roughly
constant cross section and the extrapolated magnetic field with a
strong expansion is still an open issue. In this paper, we
investigated an ultra-long but quite thin coronal loop, which could
be explained by a thin twisted flux tube model. The paper is
organized as following: \secref{sec:method} introduces the data
reduction and methods; and \secref{sec:loop} describes properties of
the coronal loop of interest; the conclusion and discussion are
presented in \secref{sec:con}.

\section{Data reduction and methods}
\label{sec:method}

\subsection{Data reduction}
\begin{figure}
\centering
\includegraphics[width=\linewidth,clip=]{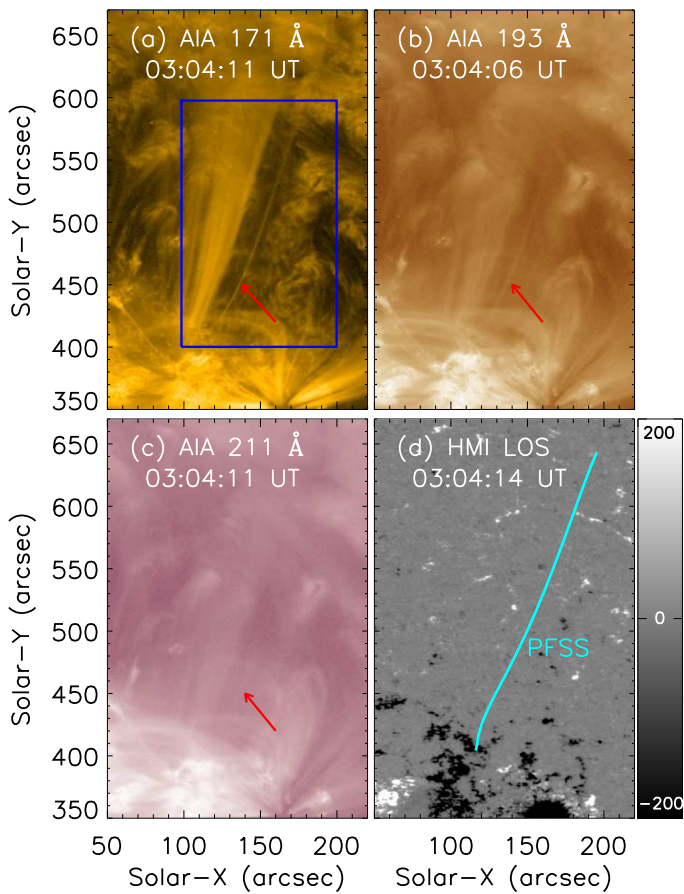}
\caption{Overview of the coronal loop on 2016 March 23.
Field-of-view observed at about 03:04~UT in AIA~171~{\AA} (a),
193~{\AA} (b), 211~{\AA} (c) and HMI LOS magnetic magnetogram (d).
The coronal loop of interest was indicated by an arrow in each
panel. Within the PFSS magnetic field extrapolation, the magnetic
field line closely aligned with this loop was overlaid (cyan curve)
with the HMI LOS magnetogram. The region of interest used in the DEM
analysis was enclosed in a blue rectangle. The evolution of this
loop is shown in a movie of anim.mp4 \label{fig:snaps}}
\end{figure}

We combined the Atmospheric Imaging Assembly \citep[AIA;][]{Lemen12}
and the Helioseismic Magnetic Imager \citep[HMI;][]{Schou12} on
onboard the Solar Dynamic Observatory \citep[SDO;][]{Pesnell12} to
observe the active region NOAA~12524 near solar disk center (N20W04)
on 2016 March 23. A unique coronal loop was predominately observed
in the AIA 171 \AA{} channel, it is also vaguely simultaneously
detectable in the AIA 193 \AA{} and 211 \AA{} channels, as shown in
\figref{fig:snaps}~(a)$-$(c). This loop had an ultra long length and
a very narrow width, which did not expand much radially along its
length. Moreover, this coronal loop retained this form for about 90
minutes, as can be seen in the movie of anim.mp4. We then used the
Potential-Field-Source-Surface (PFSS) model \citep{Schrijver03} to
extrapolate the local magnetic field line. A cyan curve in
\figref{fig:snaps}~(d) represents an open magnetic field line
derived in the PFSS model, which is closely aligned with the coronal
loop of interest.

The SDO/AIA images used in this observation has a cadence of about
12 seconds, each pixel corresponds to about 0.6\arcsec. The SDO/HMI
observes the full-disk photospheric magnetic fields. Both the AIA
images and HMI magnetograms were calibrated with the standard
routines in the Solar SoftWare package \citep{Lemen12,Schou12}.

\subsection{Loop Geometry}
This coronal loop was predominantly detectable in the AIA~171~{\AA}
channel, therefore we used the AIA~171~{\AA} images to obtain its
geometry. We created a two-dimension curvilinear coordinate,
one curve coordinate is chosen to be aligned with the spine of the
coronal loop, the second coordinate is set to be normal to the
coronal loop. We made a bilinear interpolation of the emission
intensity in the AIA~171~{\AA} channel into the curvilinear
coordinate. Then each intensity profile across the loop was fitted
with a Gaussian function plus a linear background. In order to
improve the signal-to-noise ratio, we averaged three neighboring
profiles  before fitting. The full width at half maximum (FWHM) of
that Gaussian function was considered as the loop width ($w$). The
fitting error was used as uncertainties of the loop width
\citep[e.g.,][]{Aschwanden11,Gupta19}. The error for the
AIA~171~{\AA} intensity was estimated according to the method
described by \cite{Yuan12}.

\begin{figure}
\centering
\includegraphics[width=\linewidth,clip=]{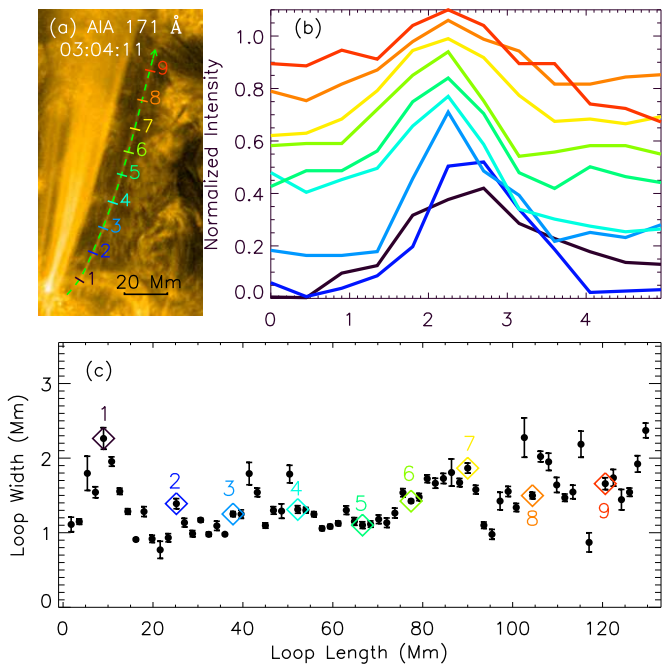}
\caption{Estimation of the loop width. (a) Smaller FOV
($\sim$76~Mm~$\times$~149~Mm) of AIA~171~{\AA} image. The loop is
highlighted by a green arrow. Nine sample cuts was marked by short
lines and numbered from 1 to 9. (b) Intensity profiles along the
cuts indicated in (a), which are normalized by their maximum
intensity, respectively. The color used in each curve is the same as
used in the numbered short lines in (a). Each profile is elevated
progressively for visualization purpose. (c) Loop width variation
along the loop length, the numbers mark the nine loop
segments in panel~(a). \label{fig:width}}
\end{figure}

In \figref{fig:width}~(a), nine cross cuts were made along the
coronal loop and were plotted with the short color lines. Each
intensity profile was scaled to a proper range and stacked in
\figref{fig:width}~(b). The loop width was measured at locations
perpendicular to the loop axis from the footpoint to the
apparent top at a distance of about 130 Mm, as indicated in
\figref{fig:width}~(c). We note that, at some positions (such as the
positions close to `8' and `9'), the obtained loop widths deviated
significantly from those of their neighbors. The reason is that some
random bright patchy background contaminated those emission
intensity profiles. On the other hand, at some positions of
(i.e., `1') the footpoint, the loop of interest had overlap with
some bright closed loops, which resulted into the broad loop
widths.

\subsection{DEM analysis}
In order to obtain the thermal properties of this loop, we focused
on a smaller FOV as marked in \figref{fig:snaps} (blue rectangle)
and performed Differential emission measure (DEM) analysis.
Observations taken from six EUV channels of SDO/AIA (94~{\AA},
131~{\AA}, 171~{\AA}, 193~{\AA}, 211~{\AA}, and 335~{\AA}) were used
to calculate the DEM(T) distribution for each pixel. We used an
improved version \citep{Su18} of the sparse inversion code
\citep{Cheung15}. The derived solutions could provide valuable
information by mapping the thermal plasma from 0.3 to 30 MK. The DEM
uncertainties were estimated from Monte Carlo (MC) simulation
\citep{Su18}. Random noise of the observed emission intensity was
added to the MC simulation and the inversion was repeated for 100
times, then the standard deviations of the 100 MC simulations were
used as the uncertainties of DEM solutions.

\begin{figure}
\centering
\includegraphics[width=\linewidth,clip=]{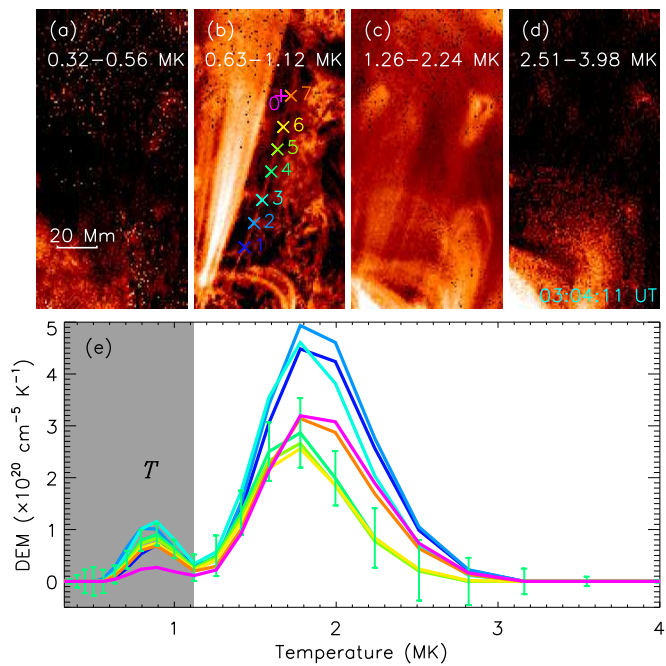}
\caption{DEM results of the target coronal loop. (a)$-$(d) Narrow
band EM maps integrated in the temperature ranges of
$0.32\,\mathrm{MK}-0.56\,\mathrm {MK}$,
$0.63\,\mathrm{MK}-1.12\,\mathrm{MK}$,
$1.26\,\mathrm{MK}-2.24\,\mathrm{MK}$, and
$2.51\,\mathrm{MK}-3.98\,\mathrm{MK}$, respectively. (e) DEM
profiles at seven selected positions (`1'$-$`7') along the loop and
one location (`0') away from the loop, the color corresponds to the
positions labeled in (b). For clarity, panel (e) only draws the
error bars at the loop position `4'. The grey region indicates the
EM integrated range. \label{fig:dems} }
\end{figure}

\figref{fig:dems}~(a)$-$(d) draws the EM \citep{Cheung15,Su18} maps
from 0.32 MK to 3.98 MK, within which coronal loops are normally
detected. These EM maps are calculated from a set of six re-binned
AIA narrow-band maps with a pixel size of 1.2\arcsec, in order to
get clear view of the structures in different temperature ranges,
whose emissions are accumulated along the line of sight (LOS) into
the observed intensity. The coronal loop was clearly seen in the
temperature range of 0.63 MK to 1.12 MK (b) and to a weaker extent
in the 1.26 MK$-$2.24 MK range (c). In order to measure the
temperature of the coronal loop of interest, we then plotted the DEM
profiles (e) of seven positions (`$\times$' in panel~b) in the
coronal loop. It can be seen that the obtained DEM profiles exhibit
two peaks at about 0.8 MK and 1.8 MK, respectively. However, the
coronal loop of interest was most clearly seen in the DEM ranging
from 0.63 MK to 1.12 MK. So we suspect that the high-temperature
peak at about 1.8 MK could originate from the emission of the
diffuse background in the AIA 211 \AA{} channel. For a comparison,
we then took the DEM profile of a reference point (`0') at the
background for cross-validation, as marked by the magenta `+'
(\figref{fig:dems}b). We noticed that the DEM profile (magenta) at
the background indeed only had a prominent peak at about 1.8 MK.

The EM was calculated by integrating the DEM over temperatures,
$\mathrm{EM}=\int \mathrm{DEM}\, \mathrm{d}T$. We only use the
temperature ranges between 0.32$-$1.12~MK, this range is the
effective temperature of the coronal loop of interest, as indicated
by the grey region in \figref{fig:dems}~(e). The EM could be
considered to the product of the square of the number density of
electrons ($n_\mathrm{e}$) and LOS depth, which could be
approximated with the loop width ($w$). In this way, the electron
number density could be calculated with
$n_\mathrm{e}=\sqrt{\mathrm{EM}/w}$. Finally, a DEM-weighted mean
temperature (such as $T=\int \mathrm{DEM}\ T \mathrm{d}T/\int
\mathrm{DEM}\ \mathrm{d}T$) is used to estimate the temperature of
this coronal loop. The errors for the density and temperature were
also calculated from the 100 MC simulations. These steps were done
for every pixel along the loop length. Then, using the
obtained number density, plasma temperature and magnetic field, we
calculated the plasma beta ($\beta$) along the coronal loop.

\section{Properties of the coronal loop}
\label{sec:loop}

\subsection{Geometry of the loop}
The coronal loop under study was very thin and ultra long. It was
detectable for a projected length of about 130 Mm. We note this as
the lower limit, since it become diffuse and invisible in the
background. The loop width was about 1 Mm at the footpoint and
expands to about 1.5 Mm to 2.0 Mm at the visible end. The loop
expansion ratio was about 1.5 to 2.0. In this dataset of about 2
hours, we observed the distinctive coronal loop to fade out
eventually, however, it had almost the constant width during its
lifetime (see the movie of anim.mp4).

In the PFSS extrapolation model, we traced a magnetic field line
that was closely aligned with the coronal loop of interest, as
indicated by a cyan curve in \figref{fig:snaps} (d). This magnetic
field line was connected to a patch of negative polarity and extends
to the outer space, therefore, this coronal loop could be regarded
as an open structure. It has an inclination in the range of
$40^{\circ}$-$80^{\circ}$ based on the estimation in the PFSS model.
The polarity at the loop footpoint has an average line-of-sight
(LOS) magnetic field component of about 100 Gauss. Along the field
line, the strength of the magnetic field is about 10 Gauss on
average, whereas the maximum field strength could reach 60 Gauss. So
we used 10 Gauss as the coronal loop's field strength.

\subsection{Thermal property of the loop}
\figref{fig:dems} presents the DEM results to the coronal loop of
interest. It is apparent that this loop is most clearly identifiable
in the EM map at 0.63~MK$-$1.12~MK (b). However, each DEM profile of
the coronal loop normally has two peaks at around 0.8~MK and 1.8~MK
(e), respectively. After a comparison with the background DEM
profile (magenta), we concluded that the DEM peak at 1.8 MK
corresponds to a strong component from the emission of the diffuse
background in the 211 {\AA} channel (Figure~\ref{fig:snaps}c).
Therefore, we estimated that the coronal loop considered here had a
temperature of about 0.8~MK.

\begin{figure}
\centering
\includegraphics[width=\linewidth,clip=]{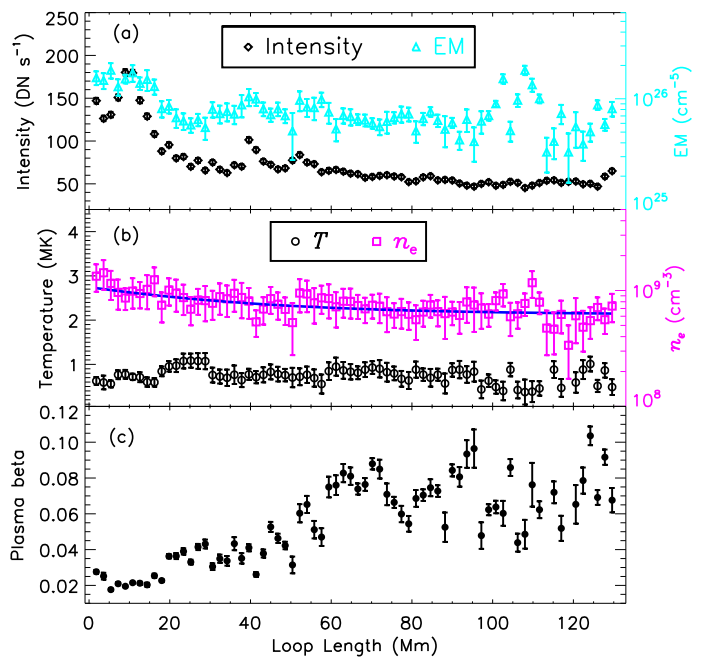}
\caption{Quantitative estimation of the coronal loop parameters. The
EM (a), the emission intensity in the AIA 171 {\AA} bandpass (a),
the plasma temperature (b), and the number density of electrons (b),
as well as the plasma beta ($\beta$) as a function of
projected length along the loop. The blue line represents the best
fitted result for the number density. \label{fig:intensity}}
\end{figure}

In Figure~\ref{fig:intensity}, we present the quantitative
estimation of the physical parameters of the coronal loop.
Figure~\ref{fig:intensity}~(a) draws the AIA 171~{\AA} intensity
\citep{Yuan12} and EM variations at the selected positions along the
loop length. They first decreased quickly with the loop length and
then became roughly stable. We notice that the AIA 171~{\AA}
intensities are much stronger at the beginning, which could be due
to the fact that some random bright patchy or closed loops
contaminated at the base of the coronal loop (see also
Figures~\ref{fig:snaps} and \ref{fig:width}).

\figref{fig:intensity}~(b) plots the plasma temperature variation
along the loop length. It shows that this coronal loop had a very
uniform temperature at $(0.7\pm0.2)\,\mathrm{MK}$, this is
consistent with our previous estimations. Since this loop was very
thin, we cannot obtain the temperature distribution across the loop.
Figure~\ref{fig:intensity}~(b) also shows the estimated number
density of electrons along the loop length, with a mean number
density of (8~$\pm$~2)$\times$10$^{8}$~cm$^{-3}$. The electron
number density was about $1.0\times10^9\,\mathrm{cm^{-3}}$ at the
footpoint, and dropped off exponentially to around
$5\times10^8\,\mathrm{cm^{-3}}$ at the end of the loop. This was a
pattern of stratification. Therefore, we fitted an exponential
function to the density profile and obtained a density scale height
of about ($38\pm13$)~Mm. This scale height only incorporate a
reduced gravity, because this corona loop is inclined with respect
to the solar radial. The theoretical density scale height is
($22.8\pm6.6$)~Mm for a plasma with a temperature of
$(0.7\pm0.2)\,\mathrm{MK}$. With the ratio of theoretical and fitted
scale heights, we estimated that the loop deviated from the gravity
vector on average by an angle of about $54^{\circ}\pm30^{\circ}$.
This value is consistent with the estimation of the extrapolated
magnetic field, i.e., $40^{\circ}-80^{\circ}$.

\figref{fig:intensity}~(c) draws the plasma beta parameter
($\beta$) as a function of the loop length. It can be seen that the
plasma beta increased from about 0.02 at the footpoint to roughly
0.1 at its visible end. The average plasma beta of this loop is
estimated to be $\sim$0.056$\pm$0.037.

\section{Conclusion and Discussion}
\label{sec:con}

In this study, we used the SDO/AIA data to observe an open coronal
loop associated with AR 12524. This coronal loops was clearly
detected in AIA 171 \AA{} and to an weaked extent in the AIA 193
\AA{} and 211 \AA{} channels. This loop was ultra-long and had a
small width, its lifetime was about 90 minutes. The loop width was
about 1.5 Mm, and a projected length about 130 Mm. The coronal loop
investigated in this study is thinner and longer than those reported
earlier. For instance, \citet{Aschwanden11} reported coronal loops
of about 2$-$4~Mm wide and 10$-$40~Mm long. Moreover, the most loops
has a lifetime of about 20$-$30~minutes \citep[e.g.,][]{Peter12},
whereas in our case, the loop survived for over one hour. The
coronal loop had a plasma temperature of $(0.7\pm0.2)\,\mathrm{MK}$,
no signification variation of temperature was detected along the
loop. This loop is relatively cold (e.g., $<$1 MK) and is
approximately uni-thermal, this is consistent with the spectroscopic
and imaging observations
\citep[e.g.,][]{Delz03,Warren08,Tripathi09}. The electron number
density was measured at about $1.0\times10^9\,\mathrm{cm^{-3}}$, and
it dropped off exponentially to about
$0.5\times10^9\,\mathrm{cm^{-3}}$ at the visible end of the loop.
This was a pattern of stratification, and then the density scale
height was measured at roughly (38$\pm$13)~Mm. The plasma
beta increases from about 0.02 to roughly 0.1, it means that the gas
pressure decreases with height by a smaller amount than the magnetic
pressure. More observational, geometrical and physical parameters
of this loop are listed in table~\ref{table:para}.

\begin{table*}[htbp]
\addtolength{\tabcolsep}{5pt}
\renewcommand{\arraystretch}{1.3}
\centering
  \caption{Observational, geometrical and physical parameters of the analyzed coronal loop.}
  \label{table:para}
\begin{tabular}{ll}
\hline \hline
Parameter                                                  &    Value                                \\
\hline
Date of observation                                 &     2016 March 23                       \\
Active region                                       &     NOAA 12524                          \\
Projected length                                    &     $\sim$130~Mm                        \\
Loop width                                                 &     1.5$\pm$0.5~Mm                \\
Expanded factor of loop width                            &     1.5$-$2.0                     \\
Plasma temperature                       & 0.7$\pm$0.2~MK                     \\
Number density of electrons              & (8$\pm$2)$\times$10$^8$~cm$^{-3}$   \\
Plasma beta                              &  0.056$\pm$0.037                  \\
Inclination angle of the extrapolated magnetic field       & 40$^{\circ}$$-$80$^{\circ}$         \\
Inclination angle inferred from stratified plasma          & 54$^{\circ}$$\pm$30$^{\circ}$       \\
Fitted density scale height                             & 38$\pm$13~Mm                        \\
Theoretical density scale height                        & 22.8$\pm$6.6~Mm                     \\
\hline \hline
\end{tabular}
\end{table*}

It should be mentioned that the little expansion was found in many
coronal loops at SXR/EUV wavelengths since the eras of TRACE and
earlier
\citep[e.g.,][]{Klimchuk92,Klimchuk00,Watko00,Peter12,Kucera19}.
These coronal loops were closed structures and often exhibited weak
expansion from double footpoints to the loop-apex
\citep{Lopez06,Brooks07}. Although an open coronal loop of about
280~Mm long was reported by \cite{Gupta19}, but its loop width
expanded from 20~Mm at the footpoint to 80~Mm at the loop top. Here,
we investigate an ultra-long ($\sim$130~Mm) and very thin
($\sim$1.5$\pm$0.5~Mm) coronal loop, which does not appear to
undergo significant expansion as it rises above the photosphere.

A very interesting point is that the loop width expanded only by a
factor of about 1.5 to 2.0, whereas this loop had a projected length
of about three pressure scale heights. The pressure has decreased by
$95\%$ (i.e, $1-e^{-3}$), the cross-section of this loop should
expand by a factor of about 19. However, this is not supported by
the observation, therefore, to explain the nearly constant
cross-section of this coronal loop, there must be unresolved
features \citep[e.g.,][]{Klimchuk00b,Petrie08}. This steady coronal
loop has a lifetime much longer than the timescales of the radiative
cooling and thermal conduction, therefore, we used a magnetostatic
model (appendix \ref{sec:tube}) to explain the small radial
expansion of this loop. We used a thin flux tube approximation and
assumed the magnetic field has a twist term
\citep[e.g.,][]{Ferriz-Mas1989}. Only when the twist term was large
enough, then the external pressure became negative. Thus, we could
find an upper limit for the twist, the detailed derivation process
was seen in the appendix \ref{sec:twist}. We then constrained the
expansion factor to be 1.5 and obtained the solution for the
external pressure. We could see that to constrain the thin magnetic
flux tube, the twist has to be smaller than about 0.65 (see,
Figure~\ref{fig:scale}). Therefore, with a sufficiently twisted
magnetic field, this coronal loop could be constrained at magnetic
equilibrium state.

If we go beyond a magnetostatic equilibrium, this loop has to be
supplied with steady flows of mass and energy, as we did not observe
any features propagating emission intensity or brightening. This
flow must have very long lifetime or should occur intermittently
with short time scales. Evidence has collected that the footpoint of
a coronal loop could have an unresolved minority-polarity, an
miniature bipolar field emerging into a constant background field.
This configuration favors small scale magnetic reconnection
\citep{Wang2019}. Repetitive reconnections provide continuous mass
and energy to the coronal loops \citep{Chitta2018}. Unfortunately we
did not observe features to support this scenario, however we shall
note that it could be that the AIA instrument is not sensitive
enough to capture these features.

At last, we stress that the loop widths are measured as the FWHM of
the cross-sectional profiles of the coronal loop detected in the
SDO/AIA~171~{\AA} channel. One AIA pixel corresponds to 0.44~Mm
\citep{Lemen12}. A loop width of about 1.5 to 2.0~Mm spans about 4-5
pixels. This measurement is obtained by fitting a intensity profile
with about 12 pixels, this practice could reach a higher accuracy
than the pixel scale, and has been used by many other researchers
\citep[e.g.,][]{Aschwanden11,Anfinogentov13,Anfinogentov19,Reale14}.

\begin{acknowledgements}
We acknowledge the anonymous referee for his/her valuable
comments. This study is supported by NSFC under grant 11973092,
11803008, 11873095, 11790300, 11790302, 11729301, 11773079, the
Youth Fund of Jiangsu No. BK20171108 and BK20191108. The Laboratory
No. 2010DP173032. D.L. is also supported by the Specialized Research
Fund for State Key Laboratories. D.Y. is supported by the NSFC grant
11803005, 11911530690,  Shenzhen Technology Project
(JCYJ20180306172239618), and Shenzhen Science and Technology program
(group No. KQTD20180410161218820). T.V.D. is supported by the
European Research Council (ERC) under the European Union Horizon
2020 research and innovation programme (grant agreement No 724326)
and the C1 grant TRACEspace of Internal Funds KU Leuven (number
C14/19/089). Y.S. is supported by the NSFC grant U1631242,
11820101002, and the Jiangsu Innovative and Entrepreneurial Talents
Program.
\end{acknowledgements}

\clearpage
\begin{appendix}
\section{Thin flux tube model}
\label{sec:tube}

In this study, we use a magnetostatic plasma to model the coronal
loop of interest. This model accounts for plasma stratification and
a static magnetic field. In our case, the loop has a lifetime
significantly greater than the timescales of thermal conduction and
radiative cooling, we resort to a magnetostatic plasma. This loop
have large aspect ratio (ratio between the loop length and radius),
we use thin-flux tube approximation \cite[e.g.,][]{Ferriz-Mas1989}
to model a twisted coronal loop. We expand the MHD quantities for a
thin cylindrical flux tube with cylindrical coordinates
$(r,\varphi,z)$. We consider an axisymmetric magnetohydrostatic
equilibrium solution to the MHD equations. We assume the velocity
$\vec{v}=0$ and set all time- and $\varphi-$derivatives to 0. In
particular, we use the following thin flux tube expansion for the
loop parameters,
\begin{eqnarray}
    B_r(r,z)&=&rB_{r1}(z) \\
    B_\varphi(r,z)&=&rB_{\varphi 1}(z)\\
    B_z(r,z)&=&B_{z0}(z)+r^2B_{z2}(z)\\
    p(r,z)&=&p_0(z)+r^2p_2(z)\\
    \rho(r,z)&=& \rho_0(z)+r^2\rho_2(z)\\
    T(r,z)&=&T_0(z)+r^2T_2(z),
\end{eqnarray}
where $\vec{B}=(B_r,B_\varphi,B_z)$ is the magnetic field, $p$ is
the pressure, $\rho$ is the density and $T$ is the temperature.

In this study, we derive the key equations for a thin and twisted
flux. We expand the magnetohydrostatic equations in the radial
component, and balance the forces in the flux tube at a certain
distance $R(z)$ with an external pressure force
$p_\mathrm{e}(R(z))$. The latter relationship is their closure
relationship for the (otherwise) infinite system of equations. The
resulting equations for the thin and twisted flux tube are
\begin{align}
    {\cal R} T_0 p_0^\prime + gp_0&=0 \label{eq:one}\\
    B_{r1}+\frac{1}{2} B_{z0}^\prime&=0 \label{eq:two}\\
    p_2+\frac{1}{2\mu}\left(-B_{z0}B_{r1}^\prime+2B_{z0}B_{z2}+2B_{\varphi 1}^2\right)&=0 \label{eq:three}\\
    B_{z0}B_{\varphi 1}^\prime - B_{z0}^\prime B_{\varphi 1}&=0 \label{eq:four}\\
    p_2^\prime-p_0^\prime\left(\frac{p_2}{p_0}-\frac{T_2}{T_0}\right)+\frac{1}{\mu}\left(B_{r1}B_{r1}^\prime-2B_{r1}B_{z2}+B_{\varphi 1}B_{\varphi 1}^\prime\right)&=0\\
    p_0+\frac{B_{z0}^2}{2\mu}+R^2\left(p_2+\frac{B_{r1}^2+B_{\varphi 1}^2}{2\mu}+\frac{B_{z0}B_{z2}}{\mu}\right)&=p_\mathrm{e}, \label{eq:seven}
\end{align}
where primes denote the derivative with respective to $z$, $\mu$ is
the magnetic permeability, $g$ is the gravity pointing in the
negative $z$-direction, and $\cal R$ is the universal gas constant
divided by the molar mass.

Since the radial expansion is of particular importance in our
problem, let us relate it to the magnetic field variables in the
problem. The general equation for the flux surface in the
$(r,z)$-plane is
\begin{equation}
\frac{dr}{dz}=\frac{B_r}{B_z}=\frac{B_{r1}}{B_{z0}}\frac{r}{1+r^2
\frac{B_{z2}}{B_{z0}}},
\end{equation}
This is not a separable equation. However, when we take $B_{z2}=0$
as done in \cite{Vasheghani2010}, the equation becomes
\begin{equation}
\frac{dr}{dz}=\frac{B_{r1}}{B_{z0}} r,
\end{equation}
Using Eq.~\ref{eq:two}, we find as solution
\begin{equation}
 R=R^\star \sqrt{\frac{B_{z0}^\star}{B_{z0}}}.
\label{eq:flux}
\end{equation}
where we use the same $\star$ notation to indicate the value at a
reference height $z=0$. Indeed, this equation expresses the
conservation of magnetic flux in a flux tube: as the magnetic field
decreases, the radius of the flux tube must increase quadratically.

\section{Solution for an expanding loop}
\label{sec:twist}

We assume that the loop radius expands exponentially with the height
\citep[see also][]{Dudik14}:
\begin{equation}
R=R^\star \exp{(z/L)},
\end{equation}
For an expansion factor $R(z_\mathrm{top})/R^\star=\eta$, we find
$L=z_\mathrm{top}/\ln{\eta}$. In particular, we can consider
$\eta=2$ or $\eta=1.5$ for the observations and
$z_\mathrm{top}=130~\mathrm{Mm}$, then $L=186~\mathrm{Mm}$ or
$320~\mathrm{Mm}$, respectively.

From the conservation of magnetic flux (Eq.~\ref{eq:flux}), we then
find
\begin{equation}
B_{z0}=B_{z0}^\star \exp{(-2z/L)},
\end{equation}
Eq.~\ref{eq:two} can be used to find the radial component
\begin{equation}
B_{r1}=\frac{B_{z0}^\star}{L} \exp{(-2z/L)},
\end{equation}
and Eq.~\ref{eq:four} can be integrated (following by
\cite{Ferriz-Mas1989}) to determine the $\varphi$-component of the
magnetic field:
\begin{equation}
B_{\varphi 1}=B_{\varphi 1}^\star \exp{(-2z/L)},
\end{equation}
From Eq.~\ref{eq:three}, we obtain
\begin{equation}
p_2=-\frac{B_{z0}^2}{\mu}\left[\frac{1}{L^2}+\left(\frac{B_{\varphi
1}^\star}{B_{z0}^\star}\right)^2\right] \exp{(-4z/L)},
\end{equation}
This equation shows that $p_2<0$, and in particular the total
pressure ($p_0+r^2p_2$) will thus be negative beyond a critical
radius.

Let us now assume that we have a constant temperature $T_0$ without
radial variation ($T_2=0$). From Eq.~\ref{eq:one}, we obtain in this
case
\begin{equation}
p_0=p_0^\star \exp{(-z/H)}, \label{eq:H}
\end{equation}
where we define the scale height $H={\cal R}T_0/g$. For a
temperature of $T_0=0.7~\mathrm{MK}$, the scale height is about
$22.8~\mathrm{Mm}$, and this value is increased to $38\,\mathrm{Mm}$
if we reduced the gravity by projection.

\begin{figure}
\centering
\includegraphics[width=1.1\linewidth]{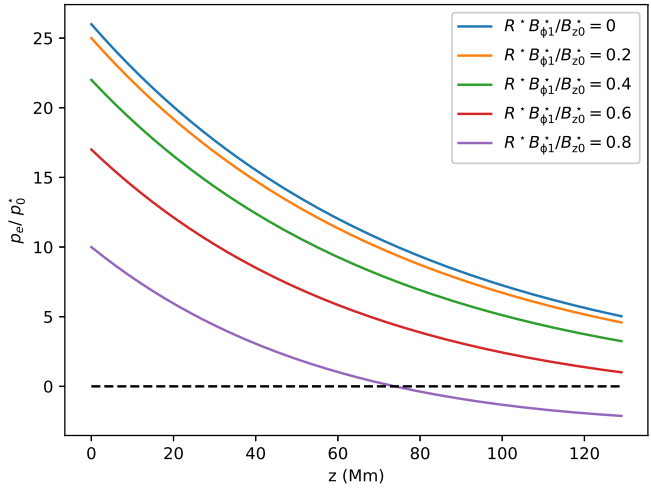}
\caption{Profiles of the gas pressure ratio of the external and
internal plasma for various twist applied. \label{fig:scale}}
\end{figure}

We then employ Eq.~\ref{eq:seven} to obtain external pressure of the
tube $p_e$:
\begin{small}
\begin{equation}
p_e(z)\!=\!p_0^\star \exp{(-z/H)}+\frac{B_{z0}^{\star
2}}{2\mu}\exp{(-4z/L)}- \frac{B_{z0}^{\star 2}}{2\mu}\exp{(-2z/L)}
\left(\frac{R^{\star 2}}{L^2}+\left(\frac{R^{\star}B_{\varphi
1}^\star}{B_{z0}^\star}\right)^2\right).
\end{equation}
\end{small}
Given that $R^\star/L$ is small, the external pressure usually
remains positive. Only when the twist term $R^\star B_{\varphi
1}^\star/B_{z0}^\star$ is large, then the external pressure becomes
negative. Thus, we can find an upper limit for the twist (see,
Figure~\ref{fig:scale}). For an expansion factor of 1.5 (or 2.0),
the upper limit of the twist is around 0.65 (or 0.5). These values
are compatible with the limits for kink instability.

\end{appendix}

\end{document}